\pdfoutput=1
\documentclass[pre,aps,twocolumn,showpacs,superscriptaddress]{revtex4-1}
\usepackage{bm}
\usepackage{mathptmx}
\usepackage{enumerate}
\usepackage{epsfig,amsopn}
\usepackage{graphicx}
\usepackage{amsmath,amssymb}
\usepackage{natbib}
\usepackage{braket}
\usepackage{comment}
\usepackage{color}
\def\bea{\begin{eqnarray}}
\def\eea{\end{eqnarray}}

\def\nn{\nonumber}

\def\la{\langle}
\def\ra{\rangle}

\newcommand{\eref}[1]{Eq.~(\ref{#1})}%
\newcommand{\fref}[1]{Fig.~\ref{#1}} %
\newcommand{\Fref}[1]{Figure~\ref{#1}}%
\newcommand{\sref}[1]{Sec.~\ref{#1}}%
\newcommand{\aref}[1]{Appendix~\ref{#1}}%
\begin{document}

\title{Driven inelastic Maxwell gas in one dimension}
\author{V. V. Prasad} 
\affiliation{The Institute of Mathematical Sciences, Taramani, Chennai - 600113, India} 
\author{Sanjib Sabhapandit} 
\affiliation{Raman Research Institute, Bangalore - 560080, India} 
\author{Abhishek Dhar} 
\affiliation{International centre for theoretical sciences, TIFR,
Bangalore - 560012, India}
\author{Onuttom Narayan}
\affiliation{University of California, Santa Cruz, California 95064, USA}
\date{\today}

\begin{abstract}
A lattice version of the driven inelastic Maxwell gas is studied in
one dimension with periodic boundary conditions. Each site $i$ of the
lattice is assigned with a scalar `velocity', $v_i$.  Nearest
neighbors on the lattice interact, with a rate $\tau_c^{-1}$,
according to an inelastic collision rule.  External driving, occurring
with a rate $\tau_w^{-1}$, sustains a steady state in the system. A
set of closed coupled equations for the evolution of the variance and
the two-point correlation is found. Steady state values of the
variance, as well as spatial correlation functions, are calculated. It
is shown exactly that the correlation function decays exponentially
with distance, and the correlation length for a large system is
determined.  Furthermore, the spatio-temporal correlation
$C(x,t)=\langle v_i(0) v_{i+x} (t)\rangle$ can also be obtained. We
find that there is an interior region $-x^* < x < x^*$, where $C(x,t)$
has a time-dependent form, whereas in the exterior region $|x| > x^*$,
the correlation function remains the same as the initial form. $C(x,t)$
exhibits second order discontinuity at the transition points $x=\pm
x^*$ and these transition points move away from the $x=0$ with a
constant speed.
  
\end{abstract}


\maketitle

\section{Introduction}
\label{introduction}

It is well-known that for a system of interacting particles in thermal equilibrium, the velocities of different particles are completely uncorrelated and the 
joint distribution of the velocities is given by the product of independent single particle Maxwell distributions. On the other hand, when a system is driven 
out-of-equilibrium, for example through application of a temperature gradient, non-zero correlations can build up between the velocities of particles \cite{garrido90}. An important class of non-equilibrium systems is {\emph{driven dissipative systems}}. An example of a dissipative system  is granular gas, which, 
in the absence of an external supply of energy, loses energy continuously due to inelastic collisions. In the presence of external driving, for example in vibrated granular systems, one can obtain non-trivial steady states \cite{Noije:98,Rouyer:00,Ben-naim:00,Santos:03,Vanzon:04,Ben-naim:05,Prasad:13}. A signature of non-equilibrium in this system is that the single-particle velocity distribution is no longer Maxwellian. 
 It is thus interesting to ask about the 
nature of correlations amongst the velocities in this system. We investigate this question in a simple lattice model of an  inelastic gas in one dimension. We calculate the exact form of the spatial correlation function of velocity for this model  in its driven steady state.

The presence of correlations in granular gases has been observed  in 
unforced~\cite{Noije:97,Baldassarri:02,Shinde:07,Jjbrey:15} as well as forced granular gases~\cite{Williams:96,Moon:01,Swift:98,vanNoije:99,Blair:01,Prevost:02,Gradenigo:11,Lasanta:15}. Different models studying unforced granular gasses
 observed power-law behavior in the spatial correlation functions~\cite{Noije:97,
Baldassarri:02,Shinde:07}.  In an early numerical study of a 
 one-dimensional granular gas, driven by uncorrelated white noise, Williams and Mackintosh 
\cite{Williams:96} observed for the density correlation function, 
a power-law behaviour when the inelasticity is large. An analytical study \cite{Swift:98} of 
a similar system of inelastic gas also  found long-range 
correlations in density and velocity in the large-$N$ limit, for finite inelasticities.
Hydrodynamic analysis of inelastic hard-sphere systems driven by white noise \cite{vanNoije:99} proposed
 correlations with logarithmic and power-law $(1/x)$ form, respectively, for two and three dimensions, which agreed with 
simulations in the near elastic regimes. In an experimental study of a granular 
gas on an inclined plane and driven by a vibrating wall at the bottom, Blair 
and Kudrolli \cite{Blair:01} also observed a power-law decay in the steady-state velocity 
correlations with the exponent ranging from $1.2$ to $2$ with decreasing system size.

 In contrast, in an experiment on a 
two-dimensional granular gas driven by a rough vibrating plane,  Prevost {\emph {et al.}} \cite{Prevost:02} found
an exponential decay in the spatial correlation of the velocities of the particles. The authors argued that the difference 
between their results and the previous ones was due to  the different driving schemes used.  In particular, the driving  in the 
analytical studies was modeled as {\emph{diffusive driving}}, with the rate 
of change of velocity due to driving equated to uncorrelated white noise. 
However, the authors in \cite{Prevost:02} argue that the driving from the wall should also be treated as inelastic 
momentum-nonconserving collisions, which suppresses long-range correlations.
To account for the different  dissipation mechanisms,  Gradenigo {\emph{ et al}.} \cite{Gradenigo:11} considered driving with a phenomenological viscous term, in addition to the white noise.  
Assuming the separation
of time-scales between the collisions and driving, they obtained
an exponential form for the velocity correlations that agreed with the experimental observations.
In the present work, 
considering a specific model of a dissipative gas, we try to understand the correlations in the case in which one does not have a time-scale separation. Also, unlike 
the previous models in which the driving is done by an Ornstein-Uhlenbeck noise 
(driving with the viscous term), we consider driving by wall-like collisions, that is motivated by the experimental
systems.

The system in which we are interested is an inelastic gas living on a one-dimensional lattice.  
In the model, a scalar velocity is ascribed to each lattice point.
 The velocities at each point change as they interact,
according to the rules of inelastic collisions.
As in one-dimensional (1D) models of granular gas with nearest-neighbor collisions, here the 
interactions are among the nearest-neighbor points on the lattice.
The model has been effective in describing the various qualitative features of cooling  1D granular gases,
such as long-range correlations and the appearance of shocks in the system~\cite{Baldassarri:02}.
The model has also been of recent interest, in developing a hydrodynamic description of granular fluids
in cooling~\cite{Lasanta:15,plata16} as well as boundary-driven steady states~\cite{Manacorda:16}. 
In the driven model presented here, in addition to the inelastic collision between nearest neighbors,  
each site has independent external driving. 

 Considering any nearest-neighbor interaction occurring with equal rates, we derive an exact set 
 of coupled equations for the evolution of the variance of the single-particle distribution 
 and the correlation functions for the system. Such a closure has been observed before, 
 for a system of Maxwell gas \cite{Prasad:14}, where spatial correlations were ignored. 
The set of equations allows one to characterize the steady-state properties for a driven system. 
For instance, the coupled relations can be used to find out whether the system goes 
to a steady state or not for various values of the parameters 
in the  driven system. 
One of our main results is the exact functional 
behavior of the spatial correlation function of the velocity field,
which shows an exponential decay at large distances. We also obtain
the spatio-temporal correlation function, and we find that it shows a second-order
discontinuity. 

Similar models have been studied before \cite{Levanony:06,Prados:11,Prados:12,Hurtado:13}~ in 
the context of granular gases as well as in the broader context of driven dissipative systems. 
In these studies, each site has an energy  instead of a momentum variable  associated with it.
Inelastic collisions are represented in the model by changing the energy of a randomly 
chosen particle to a fraction of the sum of its energy and  that of any of its 
nearest neighbors. In addition, there is dissipation and drive from a reservoir at each site or 
at the boundary. In the model considered here, one has pairwise momentum-conserving and 
energy-dissipative exchanges between neighboring particles, and it represents a somewhat 
more natural extension of the Maxwell model to incorporate spatial 
correlations ~\cite{Baldassarri:02,Lasanta:15,Manacorda:16,plata16}.

The outline of the paper is as follows. First, in \sref{Model_1d_ring} we 
introduce the model of Maxwell-like gas on a lattice with the rules of interaction 
and driving. The time evolution of the velocity distribution involves a hierarchy of 
equations as seen in the kinetic theory of granular gases. Later in \sref{equal}, 
an exact evolution of the variance and two-point correlation functions is calculated 
for the system. This helps us to characterize the time evolution of the system.
In \sref{steady state}, we  derive an exact formula for the steady-state variance 
and the equal-time correlation between the velocity variables at different sites.
Using this, one obtains an asymptotic functional form for the correlation functions 
for a large system.  We also show the extension of the above model 
where a collision between a pair occurs only when the left particle has a larger 
velocity than the right one, which mimics the real systems. 
Since this is difficult to solve analytically, we use direct simulation results to 
compare it with the model without such a constraint. As for the equal-time correlations, 
a set of equations for the spatio-temporal correlations are calculated in \sref{two time}.  We 
summarize our results in \sref{conclusion}. The details of some of the analysis are given 
in the Appendix.

\section{The model}
\label{Model_1d_ring}

We consider a one-dimensional lattice of $N$ sites ($i=1,2,\ldots,N$) with periodic 
boundary conditions ($N+i \equiv i$). Each lattice site $i$
is associated with a real scalar variable  $v_i$, which  one  calls the 
`velocity'. It should be kept in  mind that this velocity does not 
correspond to any motion in the system. 
 The system evolves in  time $t$ as follows: each  nearest-neighbor pair $(i,i+1)$ 
interacts with each other with a rate $\tau_c^{-1}$
 according to the inelastic collision rule 
\begin{equation}
\label{interparticle collision}
\begin{split}
v_{i} &= \epsilon v_i^* + (1-\epsilon) v_{i+1}^*,\\ v_{i+1}
&= (1-\epsilon) v_i^* + \epsilon v_{i+1}^*, 
\end{split} 
\end{equation}  
where, $(v^*_i,v^*_{i+1})$ and  $(v_i,v_{i+1})$  respectively are the 
pre-collision and post-collision velocities of the two interacting particles. 
Here $\epsilon=(1-r)/2$, with $r$ being the coefficient of restitution.  
For $r=1$ the collisions are elastic while $r<1$ corresponds to inelastic collisions. 
While for physical systems,  $r\in(0,1)$,  one may consider the entire range 
$r\in(-1,1)$ as a well-defined mathematical model of a dissipative gas. 

In addition to the binary inter-particle interaction, each particle is driven 
with a rate $\tau_w^{-1}$ according to 
\bea
v_i=-r_wv_i^*+\eta,
\label{particle-wall collision}
\eea
where $r_w$ is the coefficient of restitution of the wall particle collision 
with $\eta$ taken to be Gaussian noise with  variance $\sigma$ and zero mean, 
acting up on each particle independently and uncorrelated in time. 
The above driving is motivated from the collisions of the particle with a vibrating wall. 
The velocities of the particle $v_i^*$ and the vibrating wall $V_w^*$ upon collision 
 changes to new velocities $v_i$ and $V_w$ respectively  which satisfy a relation
 $(v_i-V_w)=-r_w(v_i^*-V_w^*)$. Considering a massive wall so that $V_w\approx V_w^*$,
one can obtain \eref{particle-wall collision} by substituting $(1+r_w)V_w$ by a 
random noise $\eta$. As explained before, for a Maxwell gas it is useful to extend the driving \eref{particle-wall collision} for 
negative values of  $r_w$  such that $r_w\in[-1,1]$.

Note that  $r_w=-1$ [together with the limit of $V_w\to \infty$ while keeping 
$\eta=(1+r_w) V_w$ finite] corresponds to the 
addition of Gaussian white noise~\cite{Williams:96,Noije:98}, 
which breaks the conservation of momentum  of the system, unlike the 
inelastic interparticle collisions. However, this
causes an overall diffusion of the center of mass of the system
and results in the energy of the system increasing linearly with
time~\cite{Prasad:14}. This was noted in \cite{Maynar:09}, where the authors 
add additional terms in their driving mechanism to ensure conservation of momentum. 

For $-1< r_w \leq 1$, the system reaches a non-trivial steady state~\cite{Prasad:14}. 
Note that  $ 0< r_w \leq 1$ mimics  collisions of the particle with a vibrating wall. 
The driving scheme given by \eref{particle-wall collision}, in certain limit becomes  
an Ornstein-Uhlenbeck process~\cite{Prasad:14}.

\section{Equal-time correlations}
\label{equal}
Let us define the equal time correlations $\Sigma_{i,j}(t)=\la v_i(t) v_j(t) \ra$. 
To get the equation for the time evolution of $\Sigma_{i,j}(t)$,  we follow standard procedures \cite{privman} to use 
Eqs.~(\ref{interparticle collision},\ref{particle-wall collision}) and average 
over all possible events occurring between times $t$ and $t+dt$.    In the limit $dt \to 0$ we get 
\begin{align}
\frac{d \Sigma_{i,j}}{dt}&= \left[\frac{a}{2} \Delta_2 -2b \right] \Sigma_{i,j} ~,~~{\rm for}~~|i-j| > 1 \nonumber \\
\frac{d \Sigma_{i,i+1}}{dt}&= -\left[ (1+\epsilon)a+2 b\right] \Sigma_{i,i+1} + \frac{a}{2} \left[ \Sigma_{i-1,i+1} + \Sigma_{i,i+2} \right] \nn \\
& + \frac{a\epsilon}{2} \left[ \Sigma_{i,i} + \Sigma_{i+1,i+1} \right]~, \nn \\
\frac{d \Sigma_{i,i}}{dt} &= \left[-a(1+\epsilon)-b(1-r_w)\right] \Sigma_{i,i} \nonumber \\ &+\frac{a(1-\epsilon)}{2} \left[ \Sigma_{i-1,i-1}   + \Sigma_{i+1,i+1}    \right] \nn \\ &+\epsilon a  \left[ \Sigma_{i,i-1}+\Sigma_{i,i+1} \right]+C_0~, \label{sigmaeq}
\end{align}
where  $C_{0}=\sigma^{2}/\tau_w$,
\bea
 a=2(1-\epsilon)/\tau_c~~\text{and} ~~b=(1+r_w)/\tau_w~,
\eea 
with $b,~a>0$ for  the allowed values parameters.
  In the limit of vanishing drive ($b \to 0$), these equations reduce to Eqs.~(11-14) in \cite{Lasanta:15} [after taking continuous time limit, making the identifications $r \to \alpha, L^{-1} \to \tau_c^{-1}, \Sigma_{i+k,i} \to C_k$, and  making the correction $(1-\alpha^2) \to (1-\alpha^2)/2$ in Eq.~(12) in that paper]. 
Here $\Delta_2$ is the discrete two-dimensional Laplacian operator defined by $ \Delta_2 \Sigma_{i,j}=\Sigma_{i+1,j} +\Sigma_{i-1,j} +\Sigma_{i,j+1} +\Sigma_{i,j-1} -4 \Sigma_{i,j}$.  We note that $\Sigma_{i,j}=\Sigma_{j,i}$. 
We now consider translationally invariant initial conditions such that $\Sigma_{i,j}(t)=\Sigma(|i-j|,t)$. We then get
\bea
\frac{d}{dt}Z(t)=-{\bf{A}}Z(t)+C
\label{correlation vector abstract}
\eea
where $Z(t)=[\Sigma(0,t),\Sigma(1,t),..\Sigma(n,t)]^T$, $n= N/2$ or $(N+1)/2$ respectively  for $N$ even and odd,  and the matrix 
${\bf {A}}$ is  an $(n+1)\times(n+1)$ tri-diagonal matrix of the form,

\begin{widetext}
{\footnotesize
\begin{equation}
 \bf{A} = \begin{bmatrix} 
  [2\epsilon a+b(1-r_w)]  &-2\epsilon a        &        &             &        &               &            &\\
   -\epsilon a              &[(1+\epsilon)a+2b]&-a       &             &        &\mbox{\Huge 0} &            &\\
                           &-a                  &2(a+b) &-a            &        &               &            &\\
                           &                   & \ddots & \ddots      &\ddots  &               &            &\\
                           &\mbox{\Huge{{0}}}  &        &      &-a       &2(a+b)        &-a           &\\
                           &                   &        &             &        & -2a            &2(a+b)
   \end{bmatrix}.
\label{full matrix A}
\end{equation}
}
\end{widetext}
and the column vector $C$ has $(n+1)$-dimensions 
with the only non-zero element $C_{0}=\sigma^{2}/\tau_w$. 
 The set of equations \eref{sigmaeq} can be derived alternatively 
from the BBGKY hierarchy for the distributions, as explained in \aref{Master equations}).

\begin{figure}
\includegraphics[width=.9\hsize]{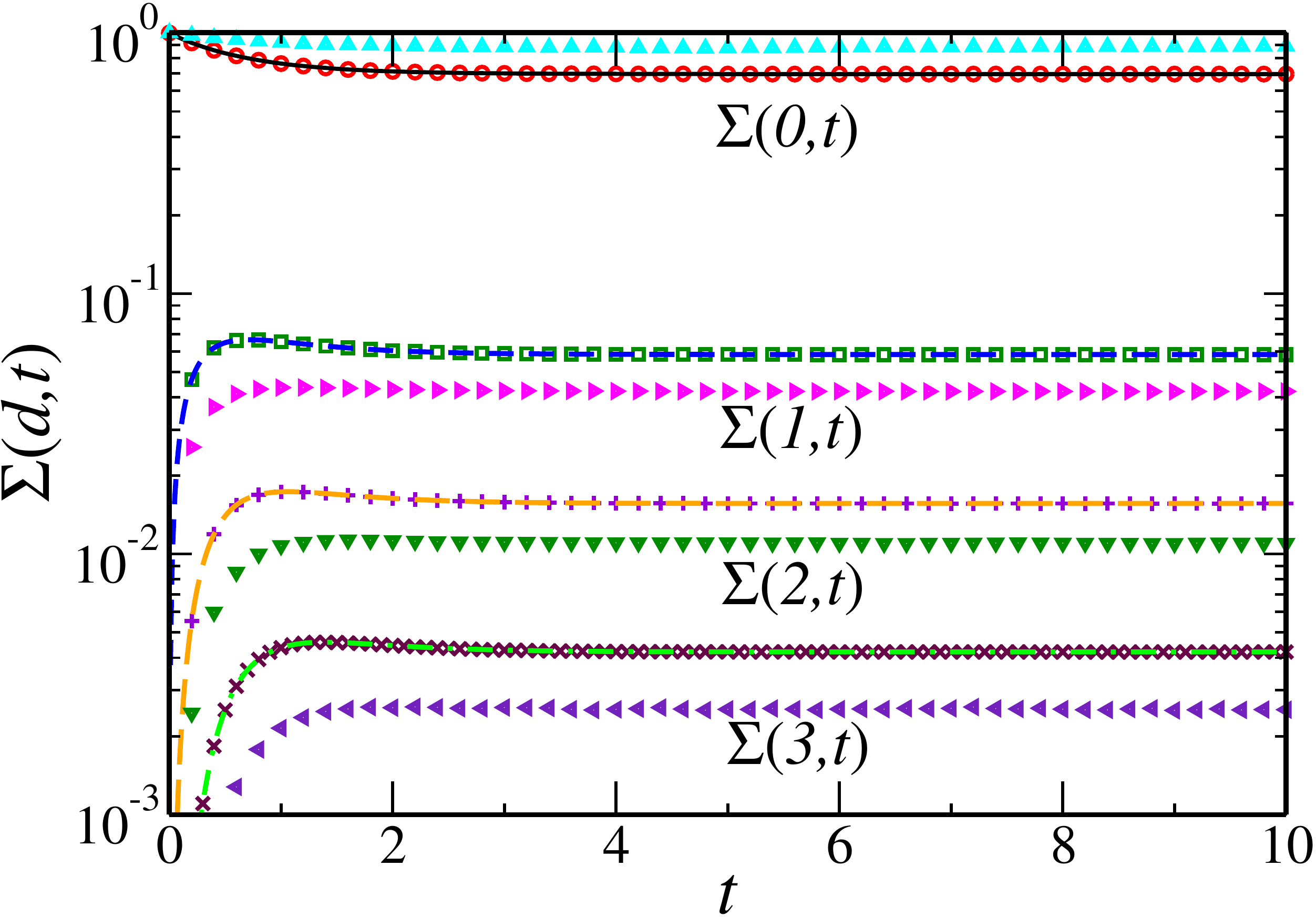}
\caption{\label{Figure1} The figure shows the evolution of $\Sigma(x,t)$ for 
x=0,1,2,3 for a 10 particle system with $r=1/2,~r_w=1/2$, $\sigma=1$, 
$\tau_c=\tau_w=1$. The triangles depict the same system with the constraint that only those pairs 
with positive relative velocity will collide.
 }
\label{sigma evolution}
\end{figure}

The evolution of $Z(t)$  can be exactly calculated from 
\eref{correlation vector abstract} which is shown in \fref{sigma evolution} along with the
numerical simulation. 
One can also consider a Maxwell gas with  the rate which depends on 
the average kinetic energy of the system. However, the steady-state properties in 
both cases follow the same statistics. 
  Further,  one  can extend the lattice model in the following way.
Instead of allowing the interaction  (\eref{interparticle collision})
to occur  with a global rate, one can consider it to occur
between the chosen nearest-neighboring pair only if their relative velocity ($v_i-v_{i+1}$), 
is positive. The condition, which is referred to as kinematic  
constraint~\cite{Baldassarri:02,Baldassarri:02a},
prevents  collision if the velocities correspond to a ``receding'' pair.
We have not been able to obtain a closed set of equations for this system. 
One can obtain the evolution of the correlations from direct simulation, and this
is plotted in \fref{sigma evolution}. One finds that the behaviour of the system 
with the kinematic constraint is different from that without the constraint.

\section{ steady state properties}
\label{steady state}
  It suffices to know the eigenvalues of $\bf A$ to see whether the 
system goes to a steady state or not. Consider the special case of 
$r_w=-1$, where the matrix has a simpler form with $b=0$. It can be 
shown that for $r_w=-1$ the determinant of the matrix $A$ vanishes, 
 and so, no steady state exists (see \aref{existence steady state rw=-1}). 
On the other hand for $r_w\ne-1$ the eigenvalues are positive (see Appendices:
 \ref{existence steady state rwne1} and \ref{existence steady state rw=1}) which 
indicates that the system goes to a steady state in this limit.

\begin{figure}
\includegraphics[width=.9\hsize]{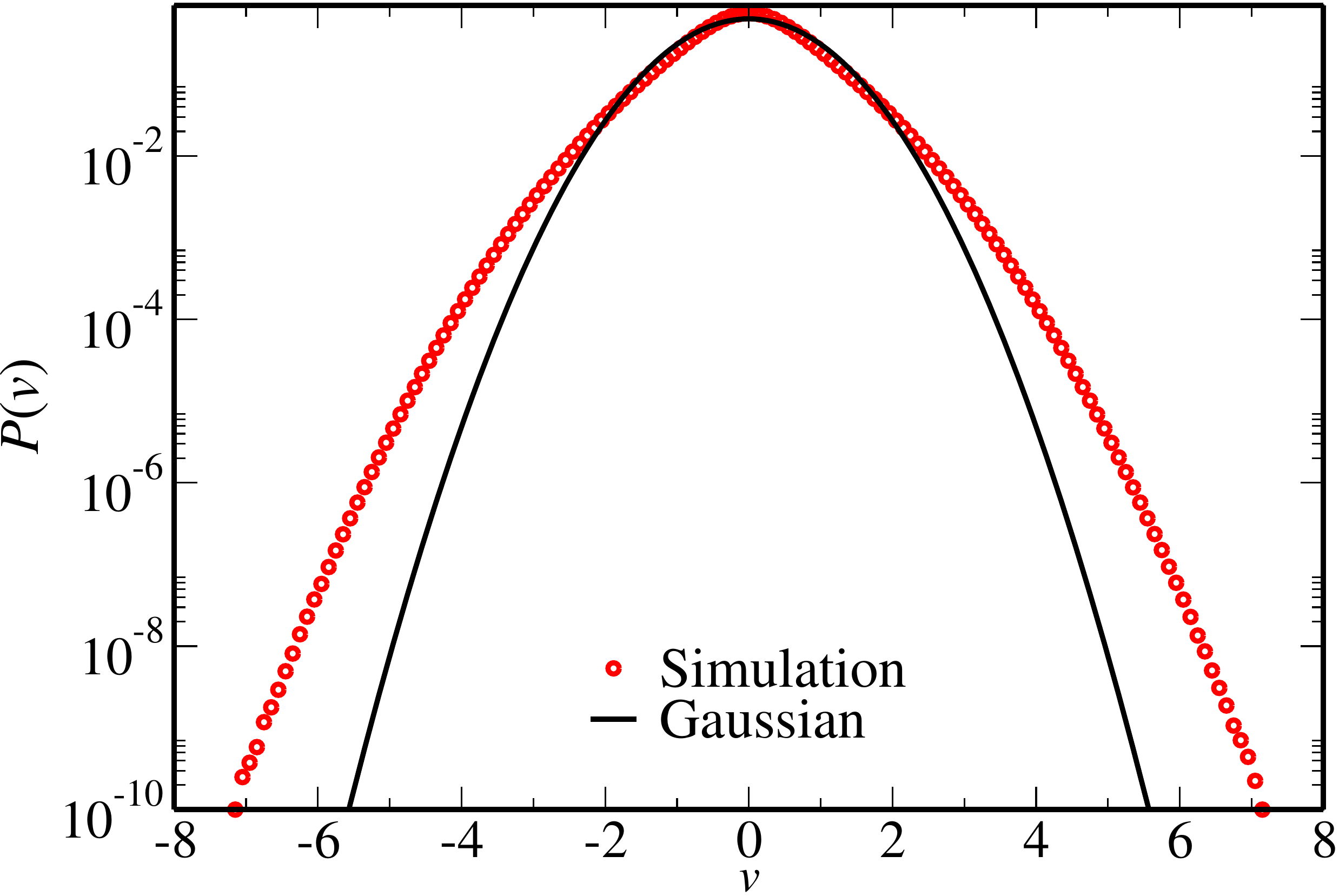}
\caption{\label{Figure1} The velocity distribution of a $50$ particle system
 with $r=1/2,~r_w=1/2$, $\sigma=1$, $\tau_c=1$ and $\tau_w=1$. The solid line shows the Gaussian with 
variance calculated for the system. One can see the deviation from Gaussian.
 }
\label{vel_pdf}
\end{figure}

\begin{figure}
\includegraphics[width=.9\hsize]{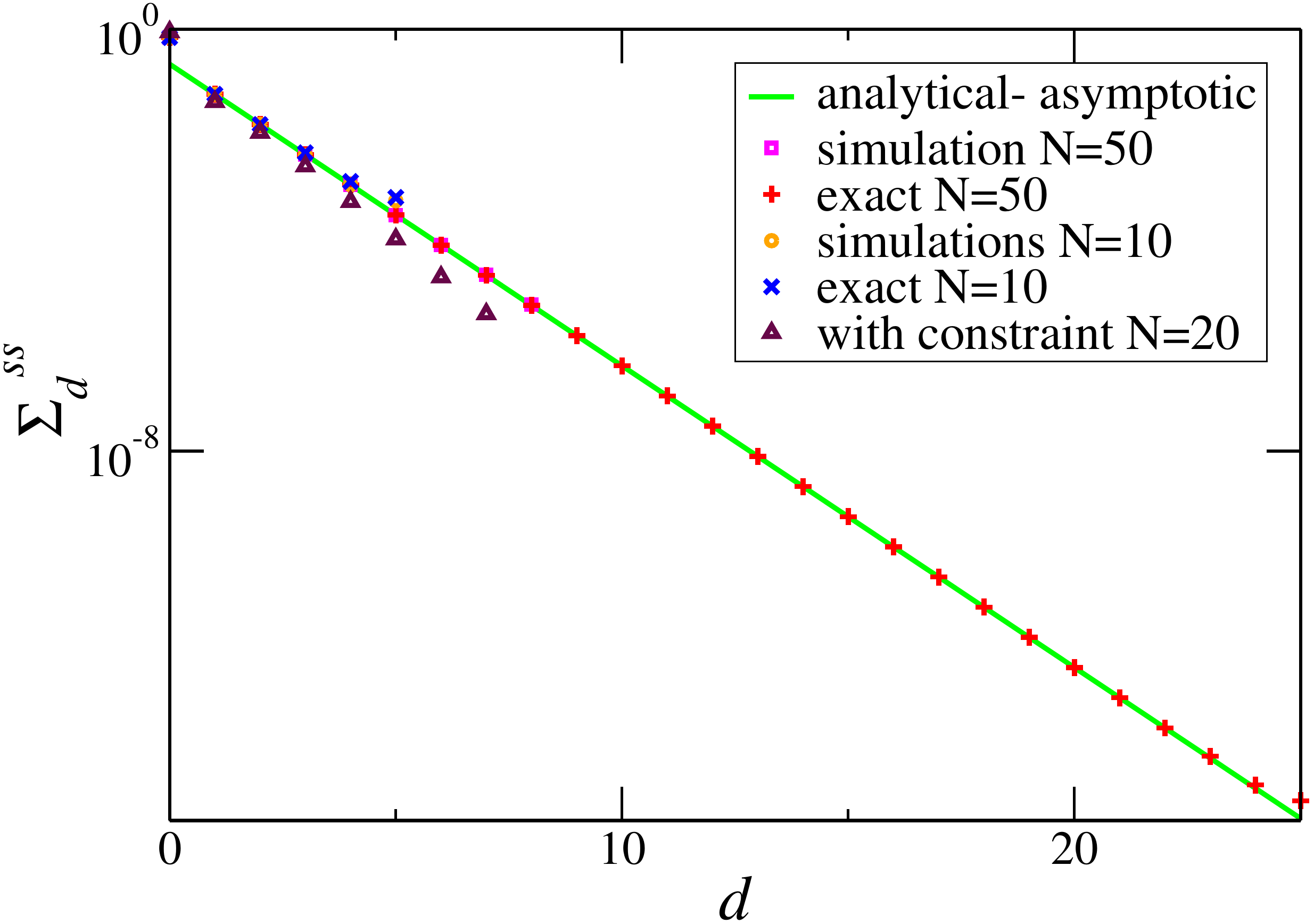}
\caption{\label{Figure2} Steady-state values of $\Sigma_{ss}(x)$  for the simulation 
of $10$ and $50$-particle systems with $r=1/2,~r_w=1/2$, $\sigma=1$, $\tau_c=1$ and 
$\tau_w=1$. The rate of collision is independent of the variance. The exact analytical results,  
given by \eref{intermediate correlation abstract},  are shown by the `$+$' symbol for  $(N=50)$
and `$\times$' for $N=10$. The asymptotic expression \eref{correlation asymptotic expression} 
is represented by the solid green line. The triangles show simulation results for the case in which 
particles collide  only when their relative velocity is positive.}
\label{sigma ss evolution}
\end{figure}

 The steady state values can be obtained by solving \eref{correlation vector abstract} 
with the left-hand side equated to zero. The elements of $Z_{ss}$, the steady-state 
correlation vector, $\Sigma_{ss}(x)=\Sigma(x,t\to\infty)$,  are obtained from,
\bea
 Z_{ss}={\bf A}^{-1}C.
\label{steady_state_abstract}
\eea
Here $x\equiv|i-j|$, denotes the separation between lattice points, which takes integer values.
Only the first column of the matrix ${\bf A}^{-1}$ suffices to calculate all 
 the elements as,
\bea
\Sigma_{ss}(x)=A^{-1}_{x0}\sigma^2/\tau_w.
\label{intermediate correlation abstract}
\eea
Calculation of $A^{-1}_{x0}$ is easy due to the tri-diagonal nature of ${\bf A}^{-1}$.
The explicit formula for $x=0$ follows as,
\begin{equation}
\begin{split}
A^{-1}_{00}=&\frac{a^n}{\det{\bf {A}}}\left\{\left[2c-(1-\epsilon)\right]\left[(s^{n-1}+s^{-(n-1)}\right]\right.
\\ &\left.-\left[s^{[n-2]}+s^{-[n-2]}\right]\right\},
\end{split}
\end{equation}
 for $x=1,2,..n$:
\begin{equation} 
\begin{split}
A^{-1}_{x0}&=\frac{\epsilon a^n}{\det {\bf {A}}}\left[s^{n-x}+s^{-(n-x)}\right]~,
\label{intermediate correlation exact}
\end{split}
\end{equation}
where
\bea  c\equiv(1+b/a)~~\text{and}~~s\equiv(c+\sqrt{c^2-1}).\eea
As $b$ and $a$ takes positive values,
$c$ and $s$ will always be greater than or equal to $1$ (equal to $1$ when $r_w=-1$).
 The determinant of the matrix ${\bf {A}}$, denoted as $\det{\bf {A}}$  has the form
\begin{equation}
\begin{split}
\det {\bf A}=&a^{n+1}\left\{K_1\left[s^{n-1}+s^{-(n-1)}\right]\right. \\
&\left.-K_2\left[s^{n-2}+s^{-(n-2)}\right]\right\},
\label{det A}
\end{split}
\end{equation}
 where $K_1, ~K_2$  are functions of ($\epsilon,c,r_w$) given by:
\begin{equation}
\begin{split}
K_1&=2\epsilon+(c-1)[4\epsilon+(1-r_w)(1+\epsilon)]+2(c-1)^2(1-r_w),\\
K_2&=2\epsilon+(1-r_w)(c-1).
\end{split}
\end{equation}
For a large system, one can calculate the asymptotic form of the correlation function $\Sigma_{ss}(x)$.
To do this, let us rearrange \eref{intermediate correlation exact} to obtain
\begin{equation} 
\begin{split}
A^{-1}_{x0}&=\frac{\epsilon a^n s^n}{\det{\bf{A}}}\left[(s^{-x}+s^{-(2n-x)})\right].
\label{intermediate correlation exact rearrnge}
\end{split}
\end{equation}
As $s>1$, in the large $n$ limit the \eref{intermediate correlation exact rearrnge} becomes,
\begin{align}
A^{-1}_{x0}&=\frac{\epsilon a^n s^n}{\det{\bf{A}}}\left[s^{-x}\right].
\label{intermediate correlation asymptote}
\end{align}
Similarly, from   \eref{det A},  for large $n$, $ \det{\bf A}$ can be shown to have the form,
\begin{equation}
\begin{split}
\det {\bf A}=a^{(n+1)}s^n\left[K_1 c s^{-1}-K_2 s^{-2}\right].
\label{det A large n}
\end{split}
\end{equation}
Thus in large $n$ limit,  $\Sigma_{ss}(x)$ has the following form:
\begin{subequations}
\begin{align}
\Sigma_d^{ss}=B \exp(-x\ln s)~,~~\\*[.2cm]
B=\frac{\epsilon s}{2(1-\epsilon)(\frac{\tau_w}{\tau_c})\left( K_1 -K_2 \right)}.
\end{align}
\label{correlation asymptotic expression}
\end{subequations}
This shows that the system has a finite correlation length
$\xi=1/\ln s$.
In \fref{sigma ss evolution} we plot the asymptotic form (\eref{correlation asymptotic expression}) 
along with the numerical (\eref{intermediate correlation abstract}) and simulation results. 
  By expanding $\ln s$ near $s=1$, one can see that the correlation length $\xi$
diverges as  $1/\sqrt{(1+r_w)}$ when $r_w$ approaches $-1$ from above. 

The  probability distribution function (PDF) of the velocity at the sites 
can be obtained from  direct simulations. In \fref{vel_pdf}
the Velocity PDF is plotted as red circles. The non-Maxwellian nature 
of the PDF is shown by comparing it with a Gaussian (black solid line)
function, which has the same variance as that of the PDF. 

 As indicated before, the above analysis cannot be 
done for a system with the kinematic constraint.
The steady-state correlation  $\Sigma_{ss}(x)$ for a system with the 
constraint is obtained from simulation and is plotted in \fref{sigma ss evolution}.
The correlation in this case is not the same as that of the model without the constraint.
 As it is difficult to obtain  $\Sigma_{ss}(x)$  for higher $x$ values from simulations,
the characteristics of the function are not clear.

\section{Two-time correlations}
\label{two time}
By proceeding as in the equal time case in Sec.~\ref{equal}, it is easy to obtain the equations of motion for the time-dependent correlation functions defined by $C_{i,j}(t)=\la v_i(t) v_j(0) \ra$, where the average is over the dynamics. 
The translation invariance of the system means that $C_{ij}(t)=C(i-j,t)$. 
We get the following equation for $C(x,t)$. 
\begin{align}
\frac{d C(x,t)}{dt} = \left[ \frac{a}{2} \Delta_1- b\right]   C(x,t)~,
\end{align}
where $\Delta_1 C(x,t) = C(x+1,t)-2C(x,t)+C(x-1,t) $~. Taking the limit $N \to \infty$ and defining  the Fourier transform 
$$\widetilde{C} (q,t)= \sum_x e^{i q x} C(x,t),$$
 we get the following solution
\begin{equation}
\widetilde{C}(q,t)= \exp \left[-(b+ a(1-\cos q)) t \right] ~\widetilde{C}(q,t=0)~,
\label{cqt}
\end{equation}
where 
\begin{align}
\widetilde{C}(q,t=0) =\sum_x e^{i q x} C(x,t=0)~.
\end{align}
From Eq.~(\ref{correlation asymptotic expression}) we have $C(x,t=0)=B \exp{(-|x|/\xi)}$, which gives
\begin{equation}
\widetilde{C}(q,t=0) = B \frac{s^2-1}{s^2+1-2s\cos{q}}~.
\end{equation}
Therefore, the two-time correlation function can be obtained as
\begin{equation}
C(x,t)=\frac{1}{2\pi}\int_{-\pi}^\pi \widetilde{C}(q,t) \, e^{-iq x}\,
dq
= B\,e^{-bt}\,C_1(x,t) , 
\label{Crt}
\end{equation}
where $C_1(x,t)$ is given by
\begin{equation}
C_1(x=\ell a t, t)=\frac{(s^2-1)}{2\pi} \int_{-\pi}^\pi 
\frac{\exp \left(-\bigl[(1-\cos q)  +iq\ell\bigr]
at\right)}{s^2+1-2s\cos{q}}\, dq.
\label{C1-integral}
\end{equation}
It immediately follows from the above integral that
$C_1(-x,t)=C_1(x,t)$. Therefore, in the following, we consider the
case $ x\ge 0$.  For large $t$, the above integral can be evaluated by
saddle point method, which suggests the form
\begin{equation}
C_1(x=\ell a t, t) \sim e^{-a tI(\ell)}.
\label{C1rt}
\end{equation}
The saddle point is given by
\begin{equation}
q^*= -i\ln \left[\ell+\sqrt{1+\ell^2}\right],
\end{equation}
which lies on the negative imaginary $q$ axis.  However, before
proceeding with the saddle-point calculation, we note that the
integrand has a simple pole on the negative imaginary $q$ axis at
$q_0=-i \ln s$ (there is also another one at $+i\ln s$ which do not
interfere with the saddle point calculation). Now, for $\ell <
(s^2-1)/(2s)$ the saddle point lies between the origin and $q_0$.
Therefore, the contour of integration can be taken through the saddle
point without crossing the pole. On the other hand, for $\ell >
(s^2-1)/(2s)$, the pole lies between the origin and the saddle
point. Therefore, in this case the dominant contribution to the
integral comes from the pole. Thus the function $I(\ell)$ is given by
\begin{equation}
I(\ell)=\begin{cases}
I_1(\ell) &\text{for}~ \ell < \ell^*\\
I_2(\ell) &\text{for}~ \ell > \ell^*
\end{cases}
\end{equation}
where $\ell^*=  (s^2-1)/(2s)$, and
\begin{align}
I_1(\ell) &= (1-\cos q^*)  +iq^*\ell\\
\label{I1}
&=
  \left(1-\sqrt{1+\ell^2} \right) +\ell \ln 
 \left[\ell+\sqrt{1+\ell^2}\right],
\intertext{and}
I_2(\ell)&= (1-\cos q_0)  +iq_0\ell\\
\label{I2}
&=- (b/a) + \ell \ln s,
\end{align}
where we have used the simplification $(s-1)^2/(2s)=(b/a)$.  It is easy
to check that $I(\ell)$ has a second order discontinuity at
$\ell=\ell^*$, that is, $I_1(\ell^*)=I_2(\ell^*)$ and
$I'_1(\ell^*)=I'_2(\ell^*)$ whereas
$I''_1(\ell^*)\not=I''_2(\ell^*)$. It is interesting to note that,
similar discontinuities of the rate function have been found recently
in various other contexts~\cite{Sabhapandit2011, Sabhapandit2012,
Pal2013, majumdar2015}.
It follows from, Eqs.~\eqref{Crt}, \eqref{C1rt}, and \eqref{I2}, that
for $|x| > \ell^* t$, we have
\begin{equation}
C(x,t) \sim B\, e^{-|x|/\xi} = C(x,t=0).
\end{equation}
Therefore, while for $|x| < \ell^* t$, the correlation function
depends on time, for $|x| > \ell^* t$, it still retains the initial
form. Such dynamical transition has been found recently in a different
context~\cite{majumdar2015}. The physical reason is that in both of these 
systems, disturbances take a finite time to propagate from one point to another.

Finally, following the method used in Ref.~\cite{Sabhapandit2012}, we
can also write down a more complete asymptotic form of $C_1(x,t)$ for
large $t$ as,
\begin{widetext}
\begin{align}
C_1(x=\ell a t, t) \approx\frac{e^{-atI_1(\ell)}}{\sqrt{2\pi a t}}
&\left[\frac{(s^2-1)}{(1+\ell^2)^{1/4}\,
\left(s^2+1-2s\sqrt{1+\ell^2}\right)}
+ \frac{\mathrm{sgn}(\ell-\ell^*)}{\sqrt{2 \bigl[I_1(\ell) -I_2(\ell)\bigr]}}
 \right]\notag\\
 + e^{-atI_2(\ell)}&\left[
\theta(\ell-\ell^*) 
-\frac{1}{2}\mathrm{sgn} (\ell-\ell^*)\,\mathrm{erfc}\sqrt{a t\bigl[I_1(\ell)-I_2(\ell) \bigr]}
\right],
\label{C1-anal}
\end{align}
\end{widetext}
where $I_1(\ell)$ and $I_2(\ell)$ are given by Eqs.~\eqref{I1}
and \eqref{I2} respectively.

\begin{figure}
\includegraphics[width=.9\hsize]{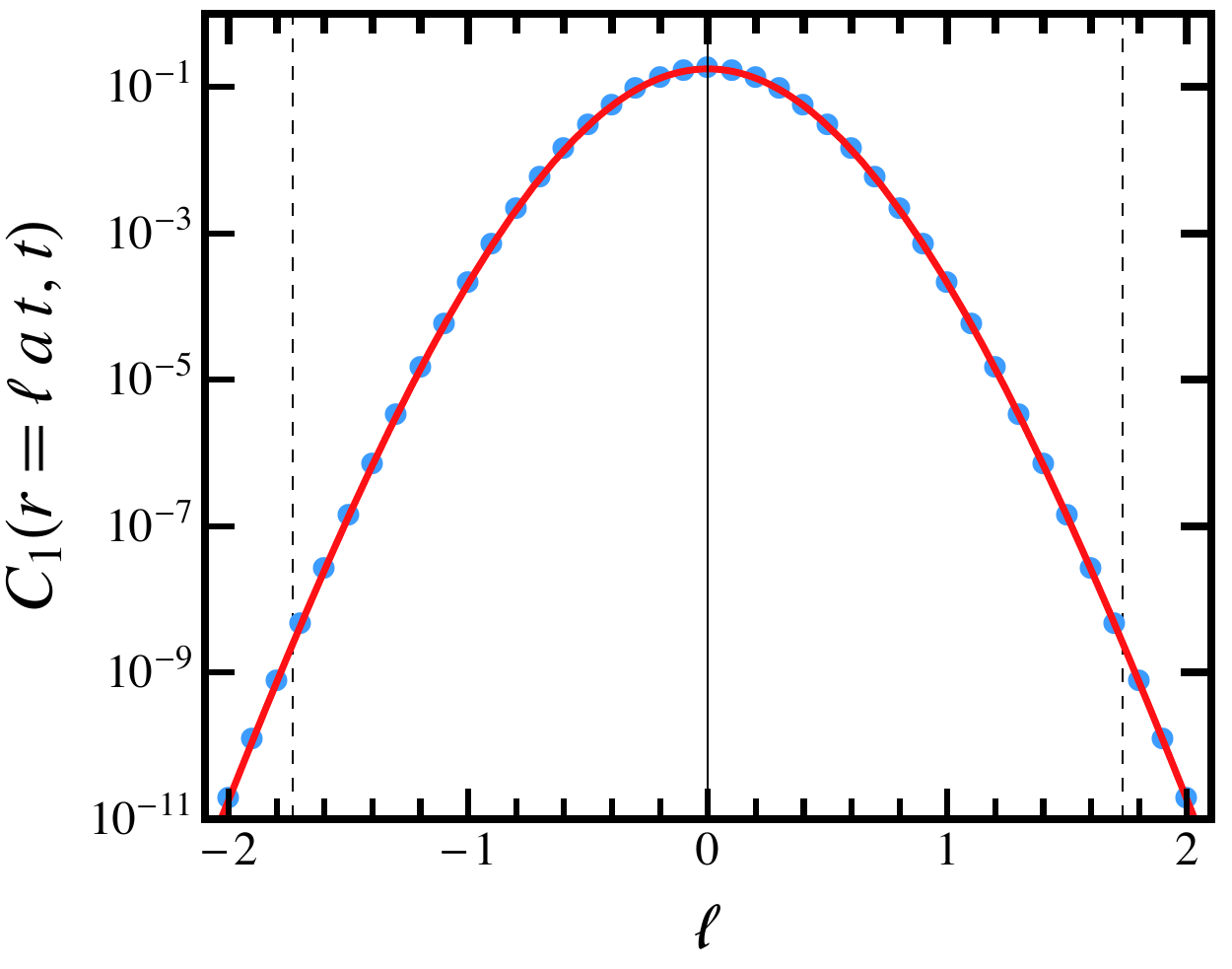}
\caption{\label{C1-fig} The points are obtained by numerically
integrating  \eref{C1-integral}, whereas the solid line represents the
analytical form given by \eref{C1-anal}. The parameters used are 
$\tau_c=\tau_w=1$, $r=r_w=1/2$ and $t=10$. These correspond to
$a=b=3/2$ and $s=2+\sqrt{3}$. The vertical dashed lines plot the
location of $\pm \ell^*$ where $\ell^*=\sqrt{3}$. }
\end{figure}

\Fref{C1-fig} compares the above result with the exact $C_1(x,t)$ obtained by
numerically integrating \eref{C1-integral} and finds perfect
agreement between the two.

As a special case, we find for large $t$ the form
\begin{equation}
C(0, t) \approx 
\frac{B(s + 1) e^{-bt}}{(s - 1)\sqrt{2\pi a t}}. 
\end{equation}
Thus there is an exponential decay as a function of time with a $1/\sqrt t$ prefactor.

\section{Conclusion}
\label{conclusion}

In this work, we studied a simple model for driven inelastic gas in
one dimension for which we find the equal-time spatial velocity
correlation functions as well as two-time correlation functions in the
steady state.  The equal-time correlations decay exponentially in
space. An interesting finding is that there exists a velocity $l^* a$
such that the decay of correlations does not propagate beyond a
distance $|x|= l^* a t$,  which leads to second order
dynamical transition in the spatio-temporal correlation function. Such
transitions have never been discussed in the context of granular
physics, and therefore, this study opens up a new direction of
research in granular physics. Hopefully, in future experiments,
such transitions could be observed in real granular systems.

 We also obtain the condition for the existence of a steady state for the
model.  Experimental studies on granular
gases driven by wall collisions, have found an exponential decay
for the spatial correlation functions of velocity~\cite{Prevost:02,Gradenigo:11}. 
Simple but exact models such as the one introduced here may facilitate a better
understanding of the observed features. It will be interesting to
study the nature of correlations in other models of granular systems with different interactions and driving mechanisms.

\begin{acknowledgments}
This research was supported in part by the International Centre for Theoretical 
Sciences (ICTS) during a visit of V.V.P. and O.N. for participating in the 
program ``Non-equilibrium statistical physics'' (Code:ICTS/Prog-NESP/2015/10).
\end{acknowledgments}

\appendix

\section{BBGKY hierarchy}
\label{Master equations}
Here we show that the equations for the correlations Eq.~(\ref{sigmaeq}) can also be derived by starting from the BBGKY hierarchy for the distribution functions.
Let $P_1(v_i,t)$ be the 1-point probability distribution function for the site 
$i$ to have the velocity variable $v_i$ at time $t$. Similarly $P_2(v_i,v_{i+x},t)$ be the 
2-site probability distribution function for the sites $i,~i+x$ to have 
velocities $v_i,~v_{i+x}$ at time $t$. Similarly defined is the 3-site probability distribution 
function $P_3(v_{i-m},v_{i},v_{i+x})$ ($\{m,x\}$ are integers less than $N$).
For the dynamics in Eqs.~(\ref{interparticle collision},\ref{particle-wall collision}), one can immediately write a set of evolution equation for the 
distributions as,
\begin{widetext}
\begin{subequations}
\label{p-eqn}
\begin{align}
\label{engy_wall_collision_point_process}
\frac{\partial}{\partial t}P_1(v_i,t)=\tau_c^{-1}\left[\int
dv_{i+1}\overline{T}(v_{i},v_{i+1})P_2(v_{i},v_{i+1},t)+ \overline{T}(v_{i-1},v_{i})P_2(v_{i-1},v_{i},t)\right]\hspace{7cm}\nn\\ +\tau_w^{-1}\left[\int dv_i^*P_1(v_i^*,t)\langle\delta\left(v_i-[-r_wv_i^*+\eta_i]\right)\rangle_{\eta_i}-P_1(v_i,t)\right],\hspace{7cm} \\
\label{corrlsn_constantini collision_point_process}
\frac{\partial}{\partial t}P_2(v_i,v_{i+x},t)=\tau_c^{-1}\left\{\overline{T}(v_i,v_{i+x})P_2(v_{i},v_{i+x},t)\delta_{x,1}+\int dv_{i-1}\overline{T}(v_{i-1},v_{i})P_3(v_{i-1},v_{i},v_{i+x},t)+\right.\hspace{6.7cm}\nn\\ \left.\left[\int dv_{i+1}\overline{T}(v_{i},v_{i+1})P_3(v_{i},v_{i+1},v_{i+x},t)+\int dv_{i+x-1}\overline{T}(v_{i+x-1},v_{i+x})P_3(v_{i},v_{i+x-1},v_{i+x},t)\right](1-\delta_{x,1})+\hspace{2.5cm}\right.\nn\\ \int dv_{i+x+1} \overline{T}(v_{i+x},v_{i+x+1})P_3(v_{i},v_{i+x},v_{i+x+1},t)\bigg\}\hspace{10cm}\nn\\
\quad+\tau_w^{-1}\left[\int dv_i^*P_2(v^*_{i},v_{i+x},t)\langle\delta\left(v_i-[-r_wv^*_{i}+\eta_i]\right)\rangle_{\eta_i}+\right.\hspace{9.6cm}\nn\\
\left.\int
dv_{i+x}^*P_2(v_{i},v^*_{i+x},t)\langle\delta\left(v_{i+x}-[-r_wv_{i+x}^*+\eta_{i+x}]\right)\rangle_{\eta_{i+x}}-2P_2(v_{i},v_{i+x},t)\right].\hspace{6.7cm}
\end{align}
\end{subequations}
\end{widetext}
and so on. Here, $\overline{T}(v_i,v_j)$ defined as,
$\overline{T}(v_i,v_j)S(v_i,v_j)= r^{-1}S(v_i^*,v_j^*)-S(v_i,v_j)$,
and acts only on the two variables designated by the arguments of the
$\overline{T}$ operator. Also $\delta_{i,j}$ is the Kronecker delta
function. The evolution of the distribution functions thus involves a
hierarchy of equations. The solution would require a closure of this
hierarchy.  As for the Maxwell particles \cite{Prasad:14}, one may ask
whether there exists such a closure in terms of the variance and two-point 
correlation functions for the one-dimensional lattice gas also.

 We calculate the evolution of the function 
$\Sigma(x,t)$, by multiplying $v_i v_{i+x}$ and integrating over $v_i$ and $v_{i+x}$.
This results in the closed set of equations for $\Sigma$ given in Eq.~(\ref{sigmaeq}).

\section{Existence of steady states for various values of $r_w$ for the inelastic gas on a 1-D lattice}
\label{existence steady state}

\subsection{Absence of steady state when $ r_w=-1$}
\label{existence steady state rw=-1}

 Here, we show that the correlation vector $Z(t)$ which evolves according to 
\eref{correlation vector abstract}, does not have a steady state when $r_w=-1$.  To show this, we observe the properties of the eigenvalues  of the matrix $\bf A$ (\eref{full matrix A}). We note that when 
$r_w=-1$, the parameter $b$ is equal to zero and the tri-diagonal matrix ${\bf A}$ has a 
simpler form (\eref{full matrix A wirh rw-1}). We denote this matrix by $ {\bf A}({r_w=-1})$.
\begin{equation}
{\footnotesize {\bf{A}}({r_w=-1}}) =\;{\mathnormal{a^{n+1}}}\;{\footnotesize \begin{bmatrix} 
     2\epsilon      &-2\epsilon          &        &            &        &               &            &\\
           -\epsilon       &(1+\epsilon)      &-1       &            &        &\mbox{\Huge 0} &            &\\
                           &-1                  &2     &-1            &        &               &            &\\
                           &                   & \ddots & \ddots     &\ddots  &               &            &\\
                           &\mbox{\Huge{{0}}}  &        &      &-1       &2            &-1           &\\
                           &                   &        &             &        &-2            &2
   \end{bmatrix}}.
\label{full matrix A wirh rw-1}
\end{equation}
 The determinant of the above $(n+1)$-th order matrix denoted as $\det {\bf A}({r_w=-1})$,  can be shown to satisfy the relation, when $n>2$:
\bea
 \det{\bf A}({r_w=-1})=2\epsilon a^{n+1} \left[\det{\bf A}^{\prime}_{n-1}- ~\det{\bf A}^{\prime}_{n-2}\right],
\label{determinant_b zero}
\eea
 where $\det{\bf A}^{\prime}_{k}$ is the determinant of  ${\bf A}^{\prime}_{k}$, which is a matrix of order $k\in \mathbb{N}$,  and has the form given below.

{\footnotesize
\begin{equation}
 \bf{A}^{\prime} = \begin{bmatrix} 
  2        &-1                  &        &        &        &               &            &\\
   -1        &2                 &-1       &        &        &\mbox{\Huge 0} &            &\\
            &-1                  &2      &-1       &        &               &            &\\
            &                   & \ddots & \ddots &\ddots  &               &            &\\
            &\mbox{\Huge{{0}}}  &        &  &-1       &2             &-1           &\\
            &                   &        &         &       &-2             &2
   \end{bmatrix} 
\label{app full matrix A prm}
\end{equation}
}
 One can find $\det\bf{A}^{\prime}_k$, as follows. Let us denote $\det{\bf{A}}^{\prime}_k\equiv D^{\prime}_k$. It can be shown to satisfy the relation, 
\bea
D^{\prime}_k-2D^{\prime}_{k-1}+D^{\prime}_{k-2}=0.
\label{d prime determinant equation}
\eea 
Using the boundary conditions, $D^{\prime}_1=2$, $D^{\prime}_2=2$, the solution of \eref{d prime determinant equation} 
can be easily obtained as, $D^{\prime}_k=\det{\bf{A}}^{\prime}_k =2$.
 Substituting this in \eref{determinant_b zero} we obtain the result, $\det{\bf A}({r_w=-1})=0$. This shows that at least one of the eigenvalue is zero, which implies the lack of steady state for the system.

\subsection{ Presence of steady state when $ |r_w|<1$}
\label{existence steady state rwne1}

Consider the matrix ${\bf A}$ (\eref{full matrix A}) when $r_w\not=-1$. We can use Gershgorin circle theorem \cite{bell:65} to predict the range of the eigenvalues of the matrix ${\bf A}$. The theorem states that any eigenvalue $\lambda$ of the matrix ${\bf A}$ should satisfy the condition:
\bea
|\lambda-{\bf A}_{ii}|\le\displaystyle\sum\limits_{j\ne i}|{\bf A}_{ij}|~~,i=0,1,2...n
\label{gresghorin}
\eea
 From the first row of ${\bf A}$, we find that:
\bea
|\lambda-\left[2\epsilon a+b(1-r_w)\right]|\le 2\epsilon a,
\eea
which says, $\lambda -b(1-r_w)\ge 0$. Similarly for $i>1$, using \eref{gresghorin} we obtain the result, $\lambda -2b\ge0$. Thus all the eigenvalues are strictly greater than zero as $b>0 $. This proves that when $|r_w|<1$,  
 the system goes to a steady state.

\subsection{Presence of steady state when $r_w=1$}
\label{existence steady state rw=1}

When $r_w=1$,  Gershgorin circle theorem provides the inequalities, $\lambda\ge 0$ from the first row of ${\bf A}({r_w=1})$ and $\lambda -2b\ge0$ from other rows of ${\bf A}({r_w=1})$, to be satisfied by the eigenvalues $\lambda$ of ${\bf A}({r_w=1})$. The above observations show that the eigenvalues of ${\bf A}({r_w=1})$ will satisfy the condition $\lambda\ge 0$. But if the system goes to a steady state, the eigenvalues should be strictly positive. This is true if the determinant, $\det{\bf A}({r_w=1})\ne0$. We show this in the following. 

As we are interested in the large system case, we consider a system with $n>2$.  
 For the system, one can show as before, that $\det{\bf A}({r_w=1})$  satisfies the equation,
\bea
\det{\bf A}({r_w=1})=2\epsilon a^{n+1}\left[(2c-1)~\det{\bf A}^{''}_{n-1}- \det{\bf A}^{''}_{n-2}\right],
\label{determinant for general A rw=1}
\eea
where ${\bf A}^{''}_{k}$ is a $k\times k$ matrix given by,

{\footnotesize
\begin{equation}
 \bf{A}^{''}_k = \begin{bmatrix} 
  2c        &-1                  &        &        &        &               &            &\\
   -1        &2c                 &-1       &        &        &\mbox{\Huge 0} &            &\\
            &-1                  &2c      &-1       &        &               &            &\\
            &                   & \ddots & \ddots &\ddots  &               &            &\\
            &\mbox{\Huge{{0}}}  &        &  &-1       &2c             &-1           &\\
            &                   &        &         &       &-2             &2c
   \end{bmatrix}. 
\label{app full matrix A rw1}
\end{equation}
}
 We define the determinant, $\det{\bf A}^{''}_{k}\equiv D^{''}_k$. From \eref{app full matrix A rw1}, one can show that
 $D^{''}_k$  satisfies the  equation,
\bea
D^{''}_{k}-2cD^{''}_{k-1}+D^{''}_{k-2}=0,~k=3,4..
\label{difference equation}
\eea
 with $c=1+b/a$.  The exact form of  $ D^{''}_k$ can be found by solving the difference equation using the initial conditions  $D^{''}_1=2c$, $D^{''}_2=4c^2-2$. The general solution for \eref{difference equation} has the form,
\bea
D^{''}_k=As^k+Bs^{-k},
\eea 
with $s=c+\sqrt{c^2-1}$. 
Using the initial conditions, the exact form of $D^{''}_k$ is found as,
\bea
D^{''}_k=s^k+s^{-k}. 
\eea 
 Substituting  $\det{\bf A}^{''}_{k}=(s^k+s^{-k})$ in \eref{determinant for general A rw=1}, one gets: 
\bea
\det{\bf A}({r_w=1})=&2\epsilon a^{(n+1)}\left((1+2b/a)\left[s^{(n-1)}+s^{-(n-1)}\right]\right.\nn \\&\left.-\left[s^{(n-2)}+s^{-(n-2)}\right]\right).
\label{determinant for general A rw=1 in terms of s}
\eea
One can rewrite the \eref{determinant for general A rw=1 in terms of s} as, 
\bea
\det{\bf A}({r_w=1})=&2\epsilon a^{(n+1)}\times \big\{\left[s^{(n-1)}-s^{(n-2)}\right.\nn\\ \left.+s^{-(n-1)}-s^{-(n-2)}\right]&+\frac{2b}{a}\left[s^{(n-1)}+s^{-(n-1)}\right] \big\}.
\label{determinant of full matrix rw=1 final}
\eea

Note that $s>1$. The material within the first set of square brackets on the right-hand 
side of \eref{determinant of full matrix rw=1 final} can be rewritten as,
\bea
\left[s^{(n-1)}-s^{(n-2)}+\frac{1}{s^{(n-1)}}-\frac{1}{s^{(n-2)}}\right]=\left(s^{2n-3}-1\right)\frac{s-1}{s^{n-1}}>0\hspace{1cm} 
\eea
 for $s>1$ and $n\ge2$.    As the term in the second set of square brackets 
 in \eref{determinant of full matrix rw=1 final}  is a positive definite quantity, 
 the right-hand side of \eref{determinant of full matrix rw=1 final} will be non-zero. 
 So the determinant of ${\bf A}({r_w=1})$ is non-zero.


\end{document}